\documentclass[journal, 10pt]{IEEEtran}

\usepackage{cite}
\usepackage{amsmath,amssymb,amsfonts}
\usepackage{algorithmic}
\usepackage{graphicx}
\usepackage{textcomp}
\usepackage[dvipsnames]{xcolor}
\usepackage{marvosym} 

\usepackage{tabularx}
\usepackage{tcolorbox}
\usepackage{caption}
\usepackage {subcaption}

\tcbuselibrary{skins}

\usepackage{multirow}
\usepackage{makecell}
\usepackage{booktabs}
\usepackage{threeparttable}
\usepackage{diagbox}

\usepackage{CJKutf8}

\usepackage{stfloats}

\def\BibTeX{{\rm B\kern-.05em{\sc i\kern-.025em b}\kern-.08em
    T\kern-.1667em\lower.7ex\hbox{E}\kern-.125emX}}

\def\eg{\textit{e.g.,}\ }

\def\etal{\textit{et al.}\ }
\def\etc{\textit{etc.}\ }

\newcommand{\cmd}[1]{\texttt{\small{#1}}}

\newcommand\blfootnote[1]{
\begingroup
\renewcommand\thefootnote{}\footnote{#1}
\addtocounter{footnote}{-1}
\endgroup
}

\begin{document}

\begin{CJK*}{UTF8}{gbsn}

\title{\fontsize{23.2pt}{25pt}\selectfont VerilogReader: LLM-Aided Hardware Test Generation}


\author{

\IEEEauthorblockN{Ruiyang Ma\textsuperscript{1}, Yuxin Yang\textsuperscript{1}, Ziqian Liu\textsuperscript{2}, Jiaxi Zhang\textsuperscript{1}, Min Li\textsuperscript{3}, Junhua Huang\textsuperscript{3}, Guojie Luo\textsuperscript{1}}

\fontsize{9.9pt}{13pt}\selectfont\textit{\textsuperscript{1}School of Computer Science, Peking University; \textsuperscript{2}School of Information, Renmin University of China; \textsuperscript{3}Noah's Ark Lab, Huawei}

\fontsize{9.1pt}{12pt}\selectfont
ruiyang@stu.pku.edu.cn, \{yxyang, zhangjiaxi, gluo\}@pku.edu.cn, liuziqian@ruc.edu.cn, minli.amoy@gmail.com, huang.hjh@outlook.com

\vspace{-2em}
}

\maketitle
\thispagestyle{empty}
\vspace{-50pt}

\begin{abstract}
Test generation has been a critical and labor-intensive process in hardware design verification. Recently, the emergence of Large Language Model (LLM) with their advanced understanding and inference capabilities, has introduced a novel approach. In this work, we investigate the integration of LLM into the Coverage Directed Test Generation (CDG) process, where the LLM functions as a Verilog Reader. It accurately grasps the code logic, thereby generating stimuli that can reach unexplored code branches. We compare our framework with random testing, using our self-designed Verilog benchmark suite. Experiments demonstrate that our framework outperforms random testing on designs within the LLM's comprehension scope. Our work also proposes prompt engineering optimizations to augment LLM's understanding scope and accuracy.

\end{abstract}


\vspace{-18pt}

\blfootnote{
This work was partly supported by the National Natural Science Foundation of China (Grant No. 62090021) and the National Key R\&D Program of China (Grant No. 2022YFB4500500). Corresponding author: Guojie Luo.

979-8-3503-7608-1/24\$31.00 \copyright2024 IEEE \hfill
}

\section{Introduction}

As hardware complexity surges, the importance of hardware verification in the development process intensifies. Undetected hardware bugs can result in substantial repercussions and considerable economic losses. To address the risk of design flaws in hardware, engineers employ two primary verification methodologies: \textit{formal verification} and \textit{dynamic verification}. 

Formal methods employs mathematical techniques to prove or disprove the correctness of a system with respect to a certain formal specification or property~\cite{formal_survey}. On the other hand, dynamic verification,  generates diverse test cases to simulate the Design Under Test (DUT), offering more flexibility and scalability than formal verification~\cite{hw_verify_survey}. Coverage targets, including code and functional coverage, serve as benchmarks for determining the thoroughness of tests. The attainment of these targets necessitates high-quality test inputs, which imposes a considerable labor burden on verification engineers.


To reduce the need for human intervention, Coverage Directed Test Generation (CDG) has emerged as a pivotal technique in automatic hardware test generation~\cite{cdg_1, cdg_2, cdg_3, rfuzz, directfuzz}. This method leverages heuristic approaches to explore the input space, with coverage states serving as basic feedback for the generation of new test cases. In situations with hard-to-reach coverpoints, supplementary circuit structural information (\eg control/data flow graph, module connectivity graph) are utilized to guide directed test generation~\cite{directfuzz, cdg_2, design2vec}.



Recently, the impressive capabilities of LLM in comprehension and inference have been highlighted.
Previous studies have shown LLM's versatility in multiple hardware tasks, such as RTL writing~\cite{llm_hw_write1, llm_hw_write2}, assertion generation~\cite{llm_hw_assert1, llm_hw_assert2} and bug fixing~\cite{llm_hw_bug_fix}. The advanced competencies of LLM present a compelling opportunity for their deployment in the field of hardware test generation. Zhang \etal have pioneered the initial step towards verifying the functional points of DUT~\cite{llm4dv}. A description of functional coverpoints is provided, following which the LLM generates input sequences. Their experimental results demonstrate a significant improvement in performance over random testing on various DUTs. Their research substantiates the capability of LLM to comprehend the high-level description of input principles and functional testpoints in the task of hardware verification.


\begin{figure}[t]
\centering
\includegraphics[width=0.45\textwidth]{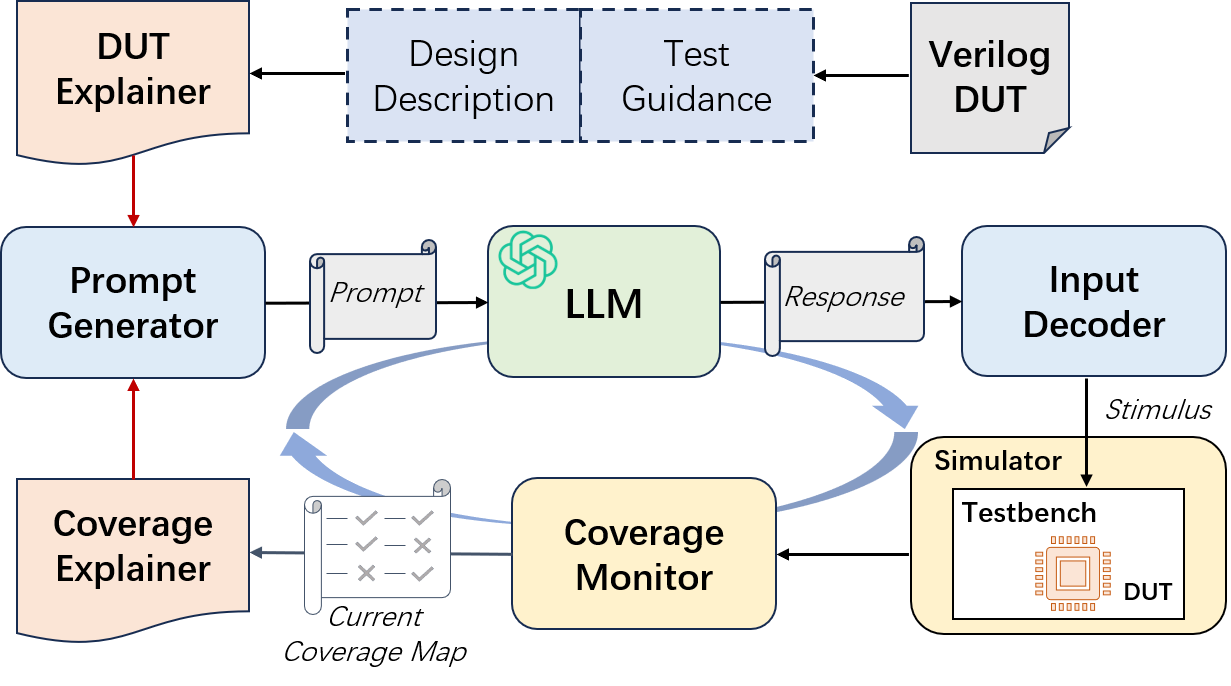}
\captionsetup{belowskip=-15pt}\caption{LLM-Aided Hardware Test Generation Workflow. }
\label{fig:workflow}
\end{figure}

While our research adopts a distinct perspective. Complementing with previous work, we have pioneered the use of LLM to specifically improve the hardware code coverage, which is a more fundamental testing target and is intrinsically linked to the Verilog code itself. This approach necessitates the shift in LLM's focus from the high-level functional testplan descriptions to the in-depth understanding of basic Verilog code logic and coverage status. That is, we repositioned the LLM as a \textit{VerilogReader}, facilitating its role as a hardware verifier to read codes and write test cases for uncovered lines or branches, consequently reducing the manual effort required for code analysis and test generation.

\blfootnote{
\textsuperscript{1}https://github.com/magicYang1573/llm-hardware-test-generation
}
In summary, our paper makes the following contributions:

\begin{itemize}
\item We open-source a framework that integrates LLM into the CDG process\textsuperscript{1}. For the first time, LLM is used as a \textit{VerilogReader} to understand Verilog code and coverage, aiming to generate tests for code coverage closure.
\item We propose \textit{Coverage Explainer} and \textit{DUT Explainer} to enrich the prompt, thereby enhancing LLM's comprehension of the design and our testing intentions. These modules also augment the extensibility of our framework.
\item We create a benchmark suite including 24 Verilog designs of simple, medium, and complex levels. Our experiments show that our framework outperforms random testing on simple- and medium-level DUTs. We also delineate the maximum Verilog reading capabilities of current LLM.
\end{itemize}




\section{Approach}

\subsection{Basic Framework}
Our study integrates LLM into the Coverage Directed Test Generation (CDG) process, as depicted in Figure~\ref{fig:workflow}. In each iteration, the LLM generates multi-cycle inputs in JSON format. These inputs are subsequently decoded by the \textit{Input Decoder} into hardware stimuli. Upon completion of simulation, the \textit{Coverage Monitor} provides current code coverage information to LLM, guiding the generation of extra input stimuli.

To generate test inputs, the LLM requires a comprehensive understanding of the Verilog DUT and the current coverage status. Given that these data are initially in non-natural-language formats, they must be transformed into a format conducive to the LLM. To this end, we have introduced two explainer modules. The \textit{Coverage Explainer} module reformats the original simulator coverage report into a more LLM-readable format, while the \textit{DUT Explainer} module enriches the DUT code with a natural language description or guidance. These modules collectively enhances the LLM's comprehension of test intentions and the DUT's functionality. Following this, the \textit{Prompt Generator} integrates these outputs to create the final prompt. 

\subsection{Prompt Generator}
\label{sec:prompt_gen}
\begin{figure}[ht]
\centering
\includegraphics[width=0.48\textwidth]{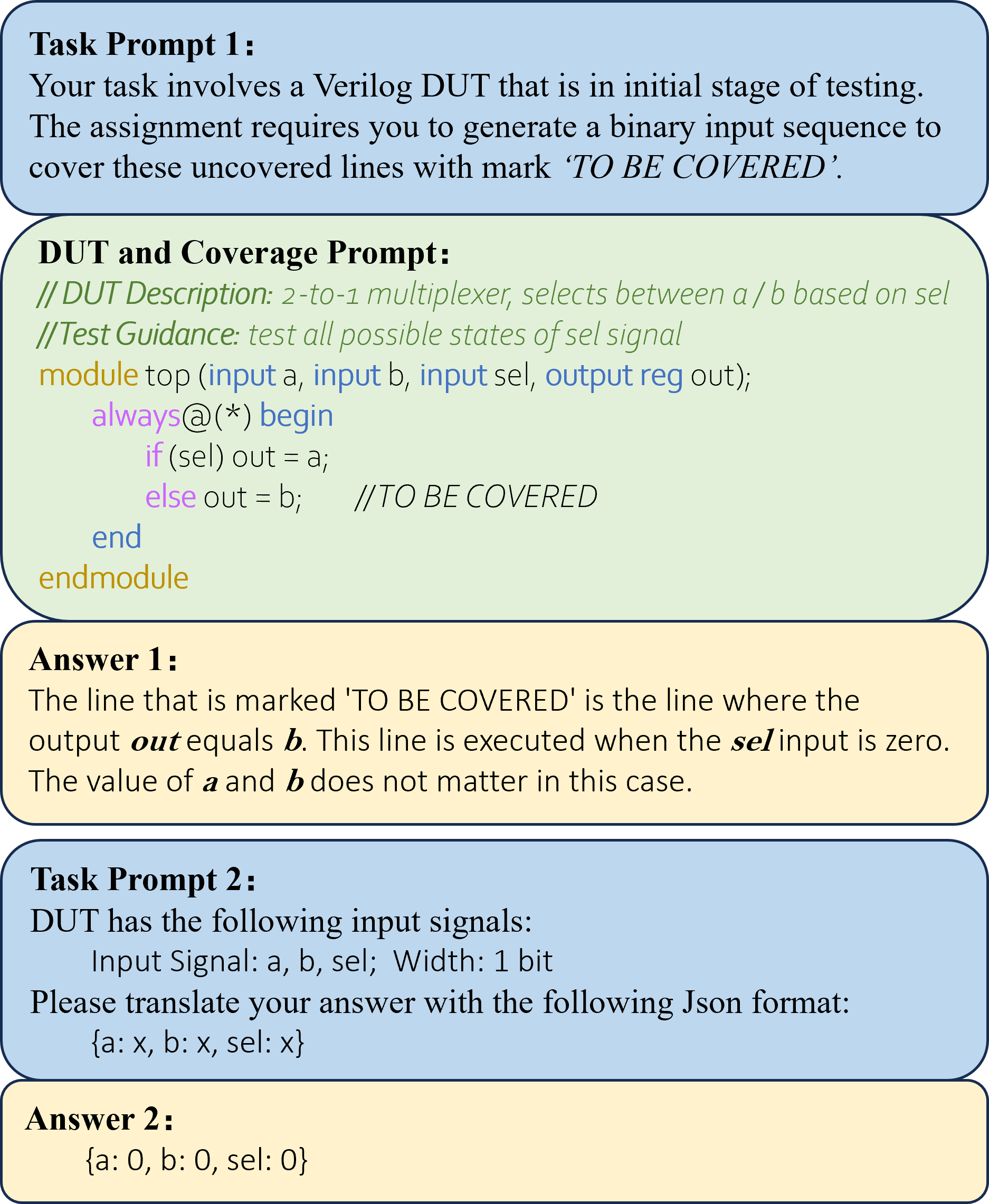}
\captionsetup{belowskip=-5pt}\caption{Example of prompts and LLM answers. }
\label{fig:qa_ex}
\end{figure}

To encourage a step-by-step thought process in the LLM, the \textit{Prompt Generator} facilitates two rounds of question-and-answer sessions in each iteration of the CDG process, thereby generating the hardware input stimulus, as depicted in Figure~\ref{fig:qa_ex}. In the first round, the LLM is informed of our objective to generate tests for unexplored code lines, incorporating details about the DUT from the \textit{DUT Explainer} and the current coverage data from the \textit{Coverage Explainer}. The LLM responds in natural language, typically mirroring its cognitive process. In the second round, we instruct the LLM to reformulate its initial response into a standardized JSON format for subsequent input decoding.

\subsection{Coverage Explainer}
\label{sec:cov_explainer}

\begin{figure}[ht]
\centering
\includegraphics[width=0.48\textwidth]{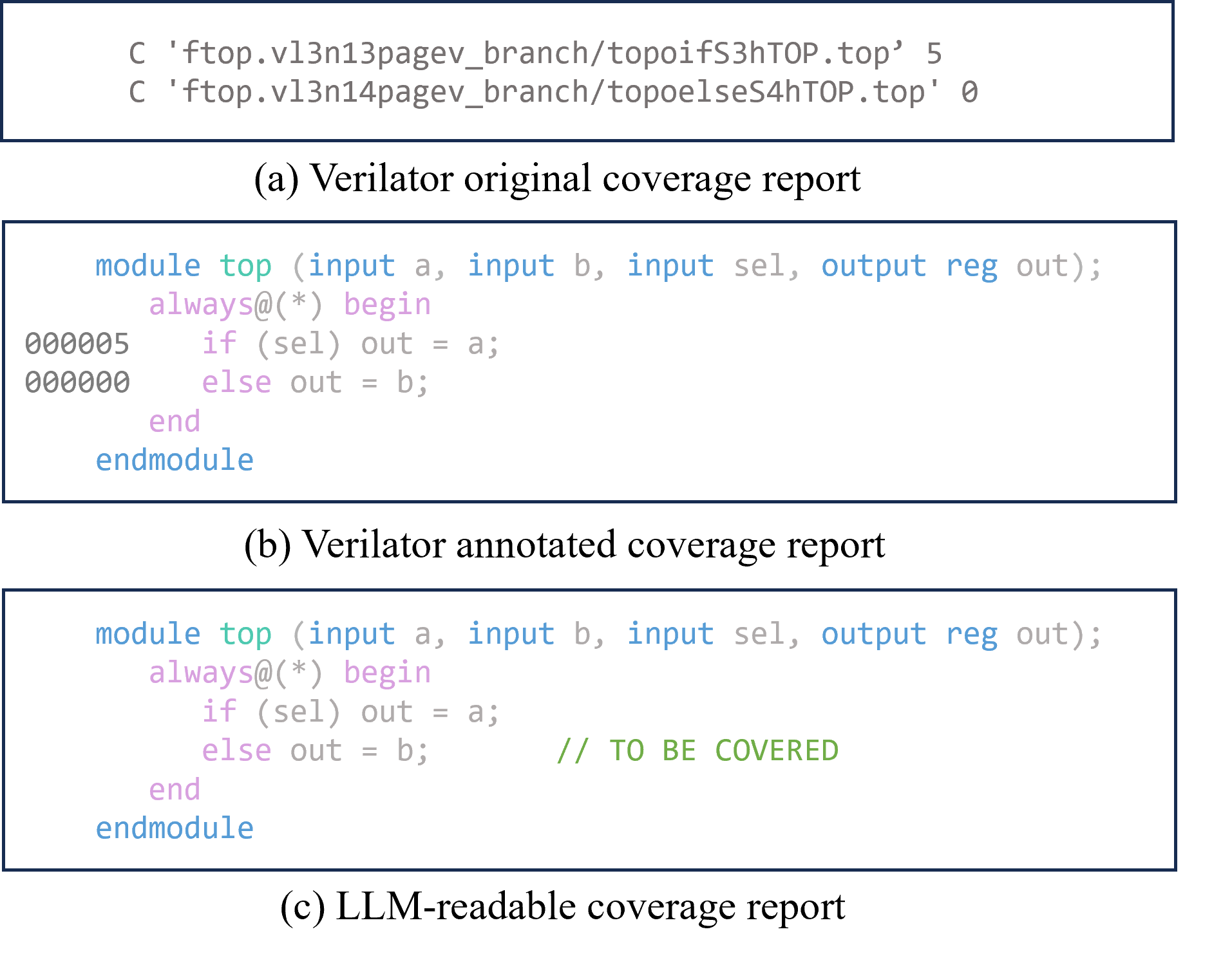}
\captionsetup{belowskip=-10pt}\caption{Comparison of three coverage report formats. }
\label{fig:cov_rpst}
\end{figure}

To enhance LLM's comprehension of the current DUT coverage, we introduce the \textit{Coverage Explainer} module, which translates the intricate coverage report into a more comprehensible format. As shown in Figure~\ref{fig:cov_rpst}(a), the original Verilator coverage report format includes each coverpoint represented by a unique identifier string and a hit count. This format is cryptic and poses readability challenges for both human users and LLM. 

A simple alternative involves using Verilator-provided \cmd{verilator\_coverage} tool to create an annotated coverage report, as depicted in Figure~\ref{fig:cov_rpst}(b). This format, which correlates coverage status with DUT source code, is more interpretable. The left-side number in each code line indicates the hit count of the line.

Despite the improvements of annotated coverage format, it still presents challenges for LLM, as LLM must identify uncovered lines, thereby increasing the complexity. To mitigate this, we suggest an advanced LLM-readable coverage report, specifically designed for our test generation task, as depicted in Figure~\ref{fig:cov_rpst}(c). This report introduces a \textit{`TO BE COVERED'} flag for lines that remain uncovered. The application of natural language to flag only the uncovered lines could facilitate a more straightforward inference process for LLM.

\subsection{DUT Explainer}
\label{sec:dut_explainer}

To augment LLM's comprehension of the DUT, we introduce the \textit{DUT Explainer} module. Given that the LLM's comprehension of Verilog code for test generation tasks is not fully optimized, this module aims to provide additional digestible information about the DUT, thereby facilitating more efficient test generation. The DUT Explainer module is designed to serve two main functions.

\textit{Design Description}, provides the LLM with a natural language explanation of the DUT's functionalities and internal logic, mitigating the LLM's incapacity to interpret Verilog code. This description can be acquired either by the LLM or manually. When acquired by the LLM, the test generation task is split into two stages: DUT understanding and input logic inference, thus alleviating LLM's workload in each phase.

\textit{Test Guidance}, enriches the LLM with supplementary information for creating tests for specific DUT. This could involve fundamental test logic rules or advice for some hard-to-cover points. For instance, when generating tests for a Finite State Machine (FSM) circuit, LLM is guided to first consider the transition to each state and then discern conditions to address any uncovered points within that state. Additionally, it could be endowed with some knowledge on reaching challenging states, thereby reducing the analytical burden on the LLM.


\section{Evaluation}

We evaluate our framework on our synthetic benchmark suite, detailed in Section~\ref{sec:benchmark}. For each design, we use Pyverilog~\cite{pyverilog} to extract input signals and automatically generate testbench interface with our framework. Verilator~\cite{verilator} serves as our simulator. The language models used in our experiments include OpenAI's GPT-4 and GPT-4-Turbo-0125~\cite{gpt4}.

\subsection{Benchmark Suite}
\label{sec:benchmark}
We created 24 Verilog designs in our benchmark suite and assigns three difficulty levels for these designs.

\subsubsection{Simple} This level involves 10 basic combinational logic circuits (\cmd{s01-s10}), such as multiplexer and ALU. The direct influence of inputs on the coverage path within the same cycle offers a straightforward inference scenario for LLM. These designs are used to assess LLM's understanding of Verilog syntax, including constructs like \cmd{always}, \cmd{case}, \cmd{assign}, \etc

\subsubsection{Medium} This level consists of 8 sequential logic circuits (\cmd{m01-m08}), such as FSMs, counters and arbiters. The coverage path of the current cycle is influenced by inputs from several preceding cycles. These designs aim to demonstrate LLM's cross-cycle inference capabilities in test generation tasks.

\subsubsection{Complex} This level encompasses 6 large-scale FSM circuits (\cmd{c01-c06}), ranging from 16 to 128 states, and two transition branches per state. It serves as a benchmark category to evaluate the upper limit of the current LLM's comprehensive ability in hardware test generation tasks.

\subsection{Comparison of Coverage Explanations}
In Section~\ref{sec:cov_explainer}, we present an LLM-readable coverage report, designed to enhance LLM's comprehension of current coverage status. To validate the utility of our coverage explanation method, we contrast it with the original and annotated coverage reports from Verilator. 

\begin{figure}[t]
\centering
\includegraphics[width=0.48\textwidth]{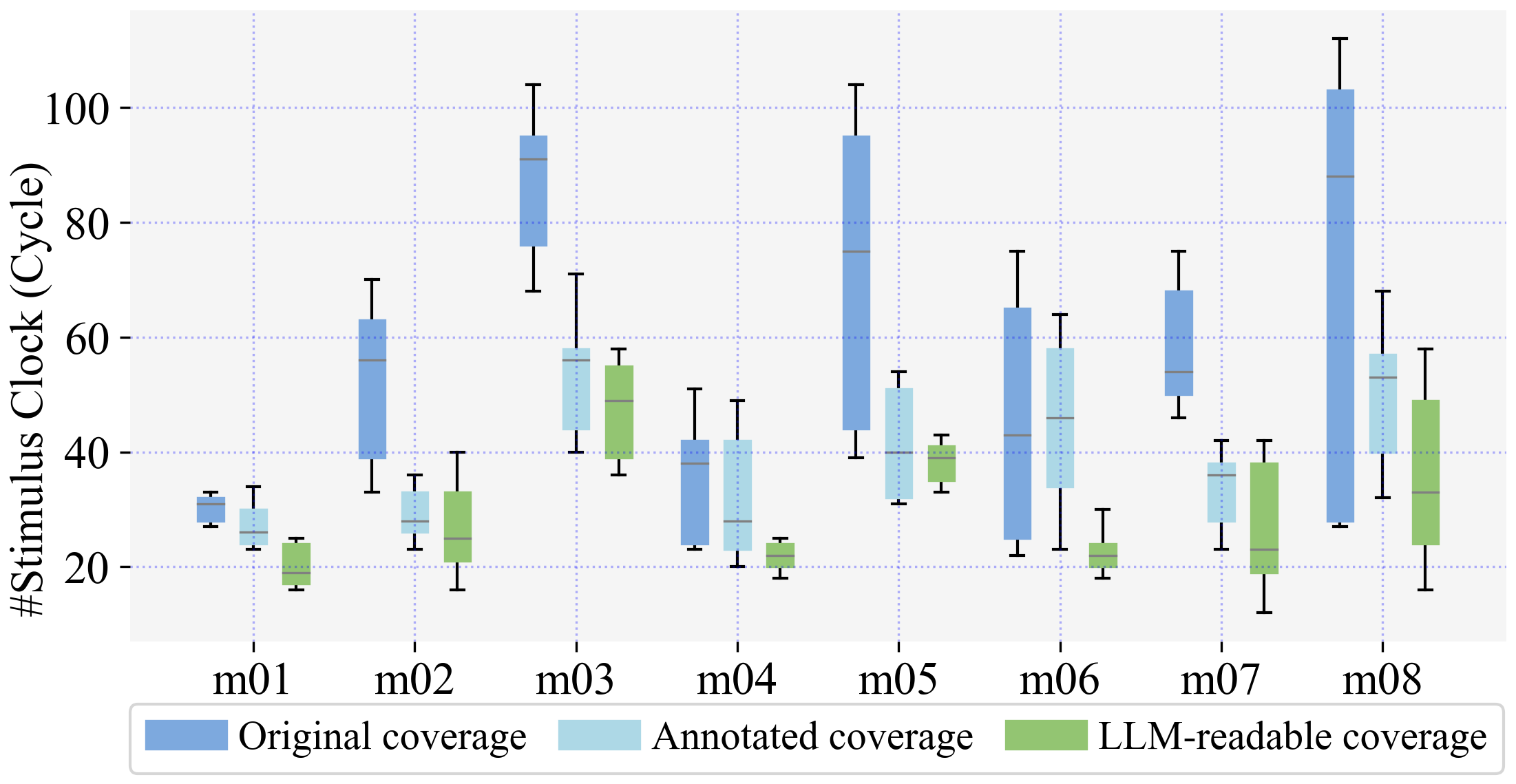}
\captionsetup{belowskip=-5pt}\caption{Comparison of coverage explanations.}
\label{fig:cov_exp}
\end{figure}

The experiments were carried out on medium-level DUTs using GPT-4 as the language model. The comparison metric was the total length of input stimulus (measured in clock cycles) required to achieve full line coverage. Given the stochastic behavior of LLM, each experiment was replicated five times. The results are represented as box (25\%ile) and whisker (75\%ile) plots, along with median lines for each DUT, as shown in Figure~\ref{fig:cov_exp}. The figure clearly indicates that the original unreadable coverage report poses the greatest challenge for LLM, whereas our LLM-readable coverage report demonstrates superior performance compared to the other two Verilator-provided reports.

\subsection{Comparison against Random Testing}

In order to evaluate the efficacy of LLM for hardware test generation, we contrast our framework with random testing. 

\begin{figure}[t]
    \centering
    \begin{subfigure}[b]{0.48\textwidth}
        \centering
        \includegraphics[width=\textwidth]{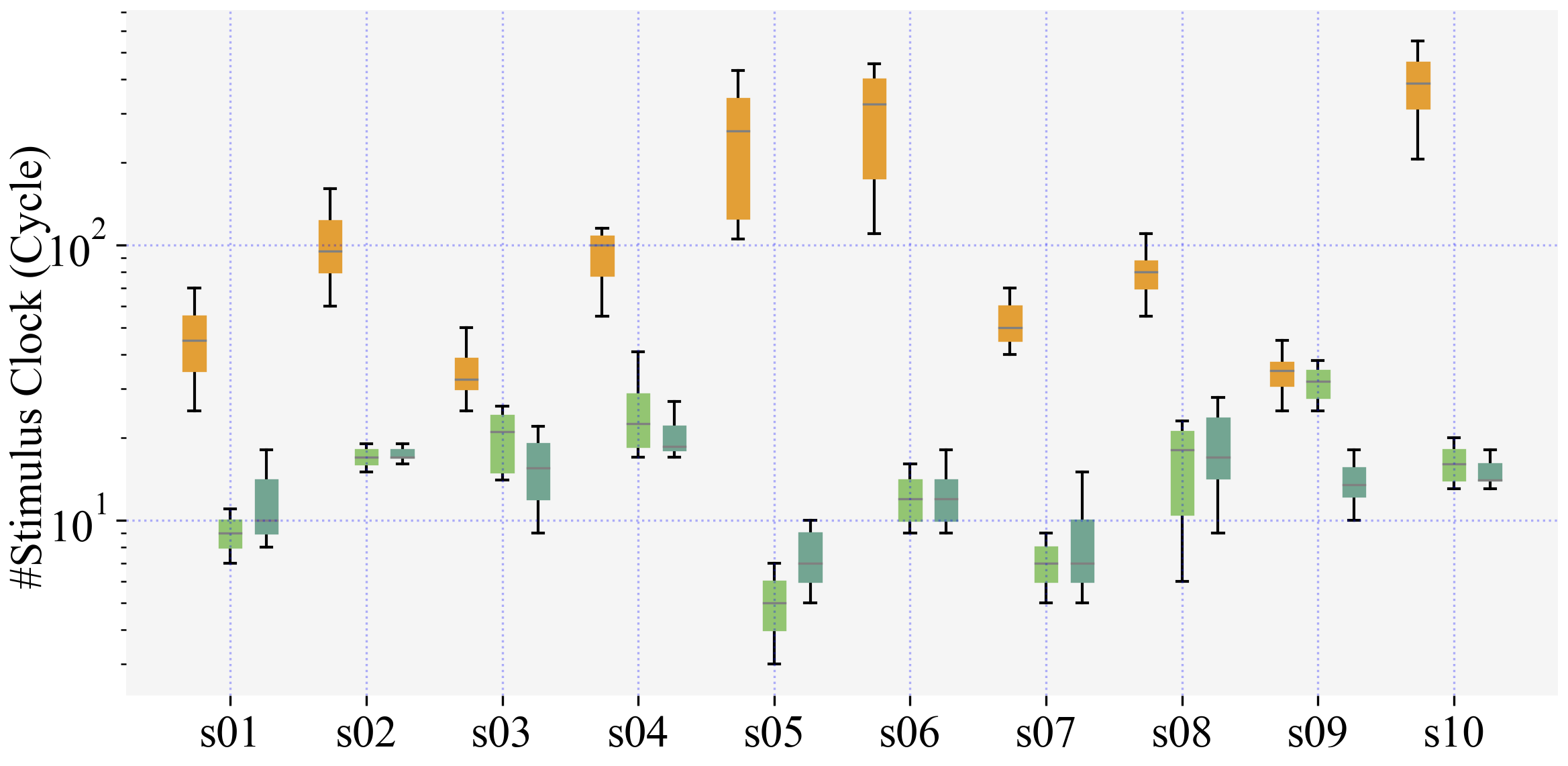}
    \end{subfigure}
    \begin{subfigure}[b]{0.48\textwidth}
        \centering
        \includegraphics[width=\textwidth]{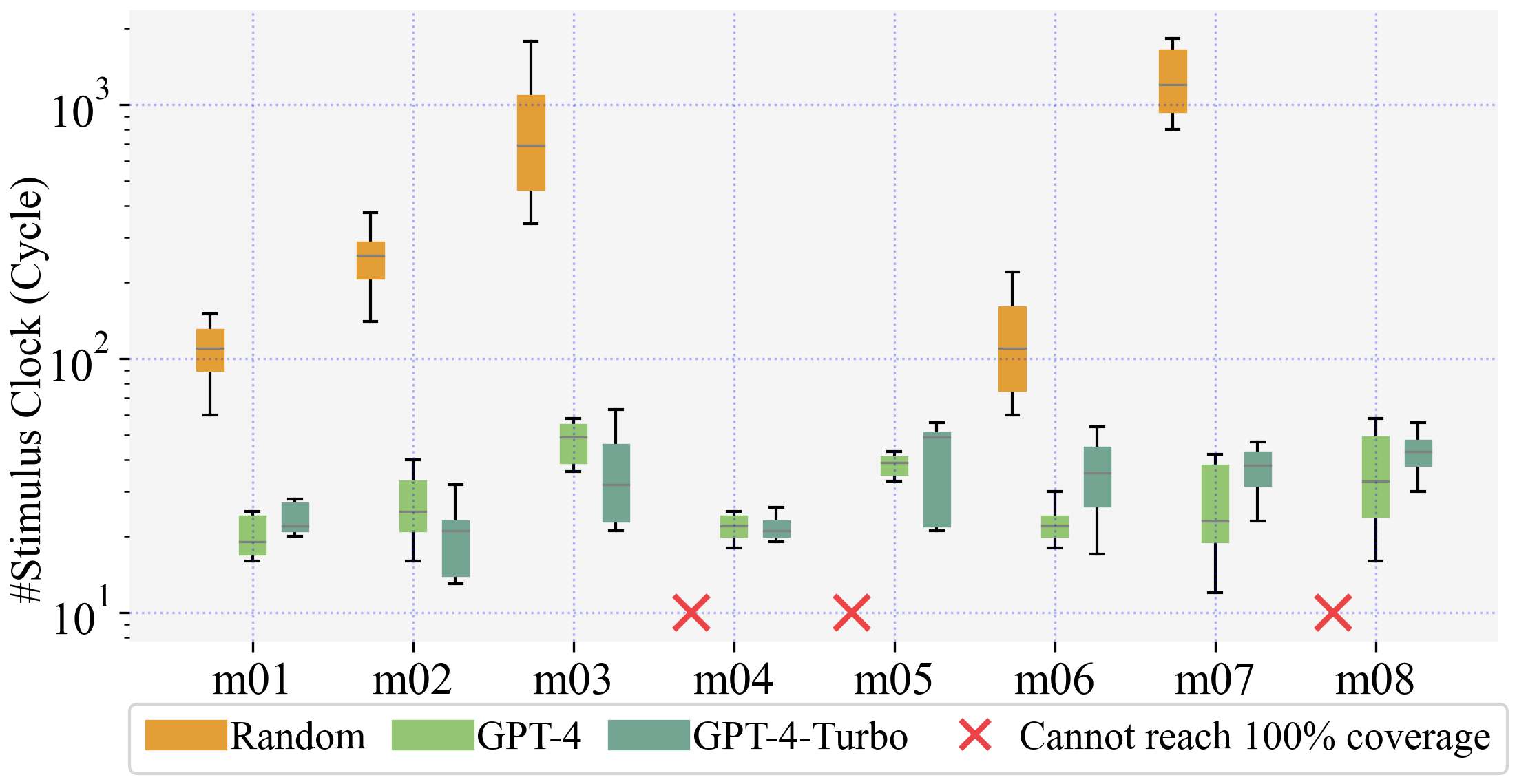}
    \end{subfigure}
    \caption{Comparison of LLM-aided test generation and random testing.}
    \label{fig:random_exp}
\end{figure}

We conducted experiments on simple- and medium-level DUTs, utilizing GPT-4 and GPT-4-Turbo as language models. We also performed five trials for each experiment. As illustrated in Figure~\ref{fig:random_exp} (log scale), LLM achieved 100\% coverage using significantly fewer inputs than random testing. The limitations of random testing became especially apparent in sequential designs with elusive branches, often failing to achieve full coverage within one-minute timeframe. In contrast, LLM could expediently reach these branches with their capacity for circuit logic analysis. Interestingly, despite GPT-4-Turbo's purported superiority, it demonstrated a similar capability to GPT-4 in hardware test generation tasks in our experiments.

\subsection{DUT Explanation Optimization}
In Section~\ref{sec:dut_explainer}, we introduce two optimization methods in \textit{DUT Explainer} module that aim to improve LLM's understanding of hardware design. Beyond providing LLM with the original Verilog code, we can supplement this with \textit{Design Description} or \textit{Test Guidance}. The former is generated by GPT-4 in our experiment, while the latter is manually written. These resources can be accessed in our open-source project.

\begin{figure}[t]
\centering
\includegraphics[width=0.48\textwidth]{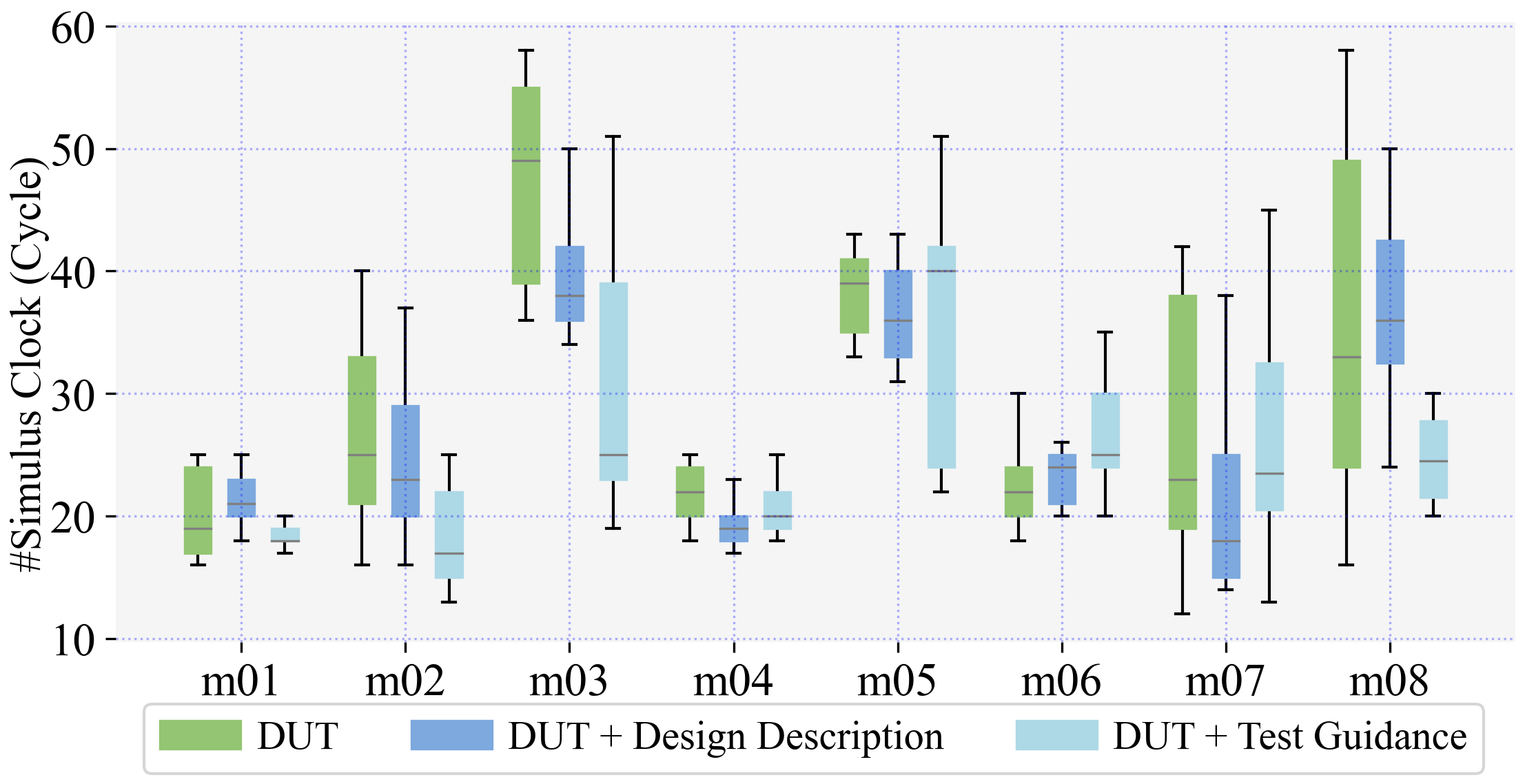}
\captionsetup{belowskip=-5pt}\caption{Effect of design description and test guidance.}
\label{fig:dut_test_guidance_exp}
\end{figure}

We carried out experiments on medium-level DUTs using GPT-4, with each experiment conducted five times. The results, presented in Figure~\ref{fig:dut_test_guidance_exp}, indicate that the inclusion of a LLM-generated \textit{Design Description} in the prompt improved LLM's understanding of the design during test generation. However, the impact of \textit{Test Guidance} was not uniformly beneficial. In designs like m05 and m06, the guidance inadvertently reduced the diversity of LLM-generated input, causing an over-reliance on our guidance and consequently stifling its capacity for self-exploration.

\subsection{LLM Reading Scalability}

In our previous experiment, we evaluated LLM's proficiency in generating tests for simple- and medium-level hardware designs, with the most complex designs consisting of around 100 lines of Verilog code. To explore the upper bounds of current LLM's capabilities for test generation, we introduced a complex level in our benchmark and employed FSMs with varying numbers of states as DUTs. Given that the largest design exceeded 500 lines of code and surpassed GPT-4's input length limitation, we chose GPT-4-Turbo for this experiment, conducting three trials and calculating the average.

\begin{figure}[t]
\centering
\includegraphics[width=0.48\textwidth]{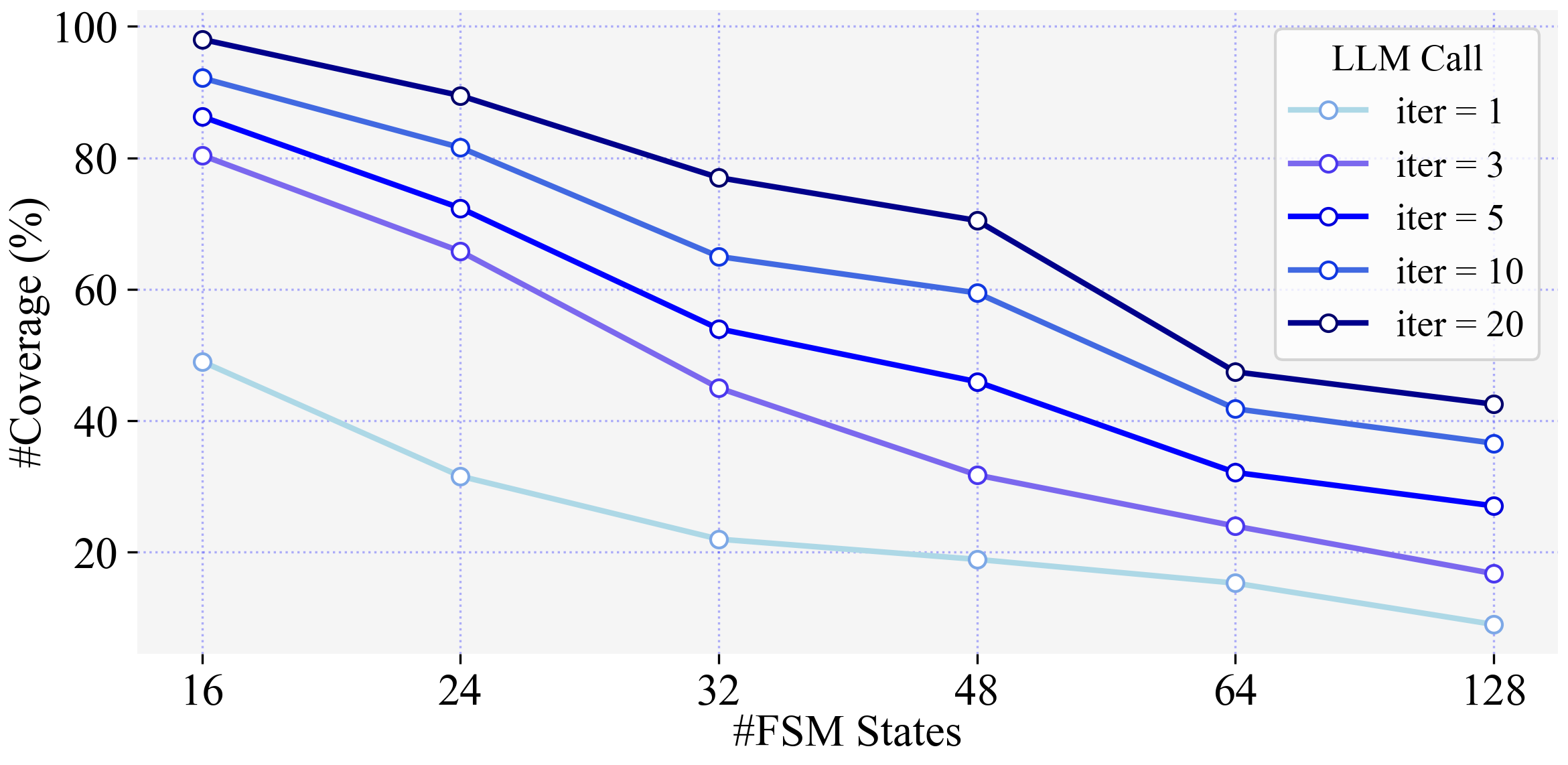}
\captionsetup{belowskip=-14pt}\caption{LLM's performance in test generation for large-scale FSMs.}
\label{fig:scale_exp}
\end{figure}

Figure~\ref{fig:scale_exp} illustrates the outcome of the experiment. It is evident that as the DUT scalability escalates, the quality of test generation precipitously declines. For an FSM with 16 states, nearly 100\% line coverage was achieved after 20 iterations of LLM calls. However, for larger FSM designs with over 64 states, the coverage cannot exceed 50\%. This reveals the LLM's inadequacies in directly processing large Verilog designs and performing intricate inferences for test generation.

\section{Discussion}

While LLM demonstrates competence in understanding simple- and medium-level DUTs, their performance diminishes with complex-level benchmarks and industry-scale hardware designs. The aspiration to employ LLM in an end-to-end manner for such designs is challenging. A substantial journey lies ahead before LLM can surpass a human hardware expert, especially in the context of Verilog code comprehension and its subsequent application in diverse EDA tasks.

One potential solution to enable the application of LLM in real-world hardware verification is to enhance our \textit{DUT Explainer}. By providing a more comprehensive high-level abstraction of the design and the verification intentions, we can guide LLM to view test generation tasks from a more macroscopic perspective. Our LLM-aided framework offers the opportunity for users to seamlessly incorporate help information during the hardware CDG process. LLM could facilitate the translation of these guidance information from natural language into actual hardware stimuli, thereby alleviating the workload of hardware verification engineers.

Furthermore, future research could focus on merging LLM with other structural AI techniques. Verilog's highly structured nature, characterized by a multitude of concurrent always blocks and module hierarchies, presents a significant challenge for LLM's decipherment. However, these structures may be more easily understood by a Graph Neural Network (GNN)~\cite{design2vec, deepgate2, llm_graph}. Therefore, the combination of LLM for regional semantic interpretation and GNN for structural interpretation could present a promising strategy to enhance the scalability of AI hardware understanding capabilities.

\section{Conclusion}
Our research primarily investigates the application of LLM in understanding Verilog designs and generating test inputs to achieve code coverage closure. We have constructed a suite of benchmarks comprising basic combinational and sequential circuits to assess our framework's efficacy. To enhance LLM's comprehension of a given DUT and the test generation task, we have introduced Coverage Explainer and DUT Explainer to enrich the prompt. Experimental results demonstrate that the LLM is capable of generating inputs and achieves full code coverage for DUTs of simple and medium complexity in our benchmarks. Future research could focus on enhancing the abstraction level of guidance information provided to LLM, or integrating LLM with GNN to capture both semantic and structural information of DUTs.

\bibliographystyle{IEEEtran}
\bibliography{long}

\end{CJK*}
\end{document}